# MiCi: A Novel Micro-Level Temporal Channel Imploration for Mobile Hosts

[1]Snehasish Kumar, [1]S. C. Rai, [2]Rajib Mall, [3]Sateesh K. Pradhan
[1]Department of Computer Science and Engineering, Silicon Institute of Technology, Bhubaneswar, India
cr.satya@gmail.com
[2]Department of Computer Science and Engineering, Indian Institute of Technology, Kharagpur, India
[3]Department of Computer Science and Application, Utkal University, Bhubaneswar, India

**Abstract** – The exponential increase of multimedia services by the mobile users requires seamless connectivity with cost-effective Quality of Service (QoS) provisioning. For providing such on-demand QoS, the network needs to utilize the radio channels among the Mobile Hosts (MHs) effectively. We use vector genetic algorithm (VGA) for temporal imploration of sharable channel(s) from the neighbouring cell(s) to fulfill the needs of a cell. We propose a new micro-level temporal channel imploration mechanism (MiCi), which promptly allocates available borrowing channel(s) of the neighbouring cell(s) to the needy cell. The novelty of MiCi is scalability, high availability, and on-demand allocation of the channels to the desired cells. The performance of our model has been tested by simulation against a standard FCA scheme as well as a Greedy Borrowing Heuristic. In all the test cases MiCi shows promising results in comparison to both the schemes.

## 1. INTRODUCTION

In a mobile cellular network each cell is assigned a set of channels to provide services to the individual MHs. The rapidly increasing demand of multimedia services with on-demand QoS provisioning to the MHs has become a major concern of mobile network designates [1]. The *Channel Assignment Problem* (CAP) concerns with allocation of available channels among the cells, so as to optimize the channel utilization. Since no adjacent cells can share any channel, CAP can be shown to be equivalent to graph-colouring problem, which is known to be NP-complete [2]. Obtaining an optimal solution for a large search space is impractical due to exponential time complexity. As a result, most of the proposed CAP in the literature is either based on some heuristics [3-5] or some evolutionary approach [6-14].

An inherent demerit of neural networks is convergence to local optima [13] and simulated annealing though guarantees global optimal solution, but suffers from a slow convergence rate [14]. GA-based approaches have been acknowledged to provide near optimal global solution with faster rate of convergence [6-11].

Several researchers have used different GA-based approaches [6-11] on macro-level channel allocation and compared their performance with Fixed Channel Allocation (FCA) [7] and some heuristic methods [6] with different network parameters. However, none of the discussed methods have so far focused on micro-level channel allocation to the cells.

Our proposed model Micro-level Channel imploring (MiCi) borrows channel(s) from the neighbouring cell(s) by temporal requesting to the cells in which sharable channels are available. It exploits the inherent potential of GA with the incorporation of seeding for initialization of the population and non-trivial single point crossover which does not lose the original information in next generation iteration process.

The CAP and various approaches are described in Section 2. Our proposed network model, Vector GA (VGA) and its use in solving the addressed problem is discussed in Section 3. The simulation environment, its parameters and performance results are presented in Section 4. Section 5 concludes this paper.

## 2. THE CHANNEL ALLOCATION PROBLEM (CAP)

The available bandwidths or channels are critical resources of a cellular network. This resource needs to be allocated among the cells so as to maximize its utilization. Any bandwidth addition to a mobile network is an expensive task [15].

The CAP is to allocate the available channels among the cells in a "fair share" manner to improve the overall performance of the network. When the number of MHs exceeds the number of available channels in a cell, the probability of call blocking, call dropping and packet loss rate increase. In such situations channel(s) may be borrowed from the neighbouring cells (if available) to fulfill the QoS demands of the users. Now the problem is to determine when and from whom, how much to borrow. Several methods like FCA (Fixed Channel Allocation) [4, 5], DCA (Dynamic Channel Allocation) [3, 4, 10, 12, 15] and HCA (Hybrid Channel Allocation) [3, 6, 7] with different performance metrics have been proposed earlier. Our work focuses on the FCA scheme with channel borrowing optimized using a Vector Genetic Algorithm (VGA) on a micro level.

An MH can move from one cell to another cell. The transition of the MH from one to its neighbouring cell may suffer blocking due to non-availability of required channel in the new cell. The objective of CAP is to minimize the average blocking probability. There are several metrics which may be used to evaluate the performance such as the average number of "*hot cells*" in the network region. A cell is termed to be hot if it does not possess a channel to allocate to a host in case of handoff.

Several methods like channel sharing and cell splitting also address this problem. Considering the type of borrowing from other cells, channel borrowing schemes can also be enumerated as macro-level or micro-level. In macro-level approach, all the cells of the network are evaluated and then a decision is taken based on the performance metrics obtained. However in a micro-level approach, the evaluation of the problem is only considered at a cellular level. That is, only a single cell along with its neighbours is considered before a borrowing decision is made.

The advantage of using a micro-level approach as compared to a macro-level approach is that in the decision making step, the size of the solution space is reduced and the rate of convergence of our proposed VGA improves. Also in a micro-approach, we only address the cells where borrowing is required, redundant solutions are not present in the final solution space. The implementation of the micro-level approach is also better suited to current network infrastructure as the borrowing decision can be made at a Base Station (BS), rather than at a centralised location such as the Mobile Switching Centre (MSC) thereby distributing the load.

*Frequency Reuse*: The micro-level channel borrowing inhibits frequency reuse among neighbouring cells. The incorporation of VGA in this scheme tries to optimize the allocation of available bandwidth in the network. Allocation of a single channel to two or more cells has two major benefits: requirement of lower power by the MHs for transreceiving and overall increase of battery life.

Channels that use the same frequencies in a cellular system are called *co-channels*. The allowable distance between co-channels is determined primarily by the transmitter power used in each cell site/BS. The minimum number of channels required *(N)* to serve the entire coverage area can be expressed as:

$$N = \frac{1}{3}\sigma^2$$, for hexagonal cells

Here $\sigma$ is defined as $\frac{D}{R_a}$, where $R_a$ is the radius of the cells and D is the physical distance between the two cell centres. N can only assume integral values *3, 4, 7, 9,...* generally represented by the series *(i+j)2-ij*, with i and j being integers. For our simulation we consider $\sigma$ = *4.45* and *N = 7*.

Due to *co*-channel interference, the same channel sets are not used in neighbouring cells. In our model, the entire channel set is divided into seven cells and assigned among a group of seven adjoining cells.

## 3. MiCi: THE PROPOSED MODEL

Our proposed model MiCi has two parts, one is Network Model and another is Vector Genetic Algorithm (VGA) model. Both the models are described in the following sub-sections.

### 3.1 Network Model

We consider the network to be of rectangular in shape. It is composed of 25 hexagonal cells. Any cell of the network may have up to 6 neighbours as shown in Figure 1. There is a marginal increase in complexity due to rectangular shaped network, which can be easily overcome by using a simple *lookup-table,* tabulated only once, at the beginning of execution of MiCi. The cells are sequentially numbered from left-to-right and top-to-bottom manner, with the first cell numbered as 1. Thus the top-left, top-right, bottom-left and bottom-right cells are numbered as 1, 5, 21 and 25 respectively. It is clear from the figure that all cells don't have equal number of neighbouring cells. The cells on the top and bottom row have four neighbours or less, the cells in the sides and corners also have a varying number of neighbours. All these information are initialized and maintained in the *lookup-table*.

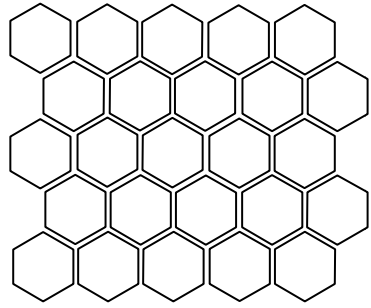

**Figure 1.** Our Network Model

*Mobile Hosts*: An MH in the simulated environment is an entity which requires a channel in order to maintain an active network transaction. The movement of the MHs is of concern during simulation of the model. Our network region is similar to a city area, where the movement of the host is random. Thus either a host moves from one cell to any one of its neighbouring cells or remains in the same cell in the next iteration. The total number of MHs is kept constant by not allowing any host to leave the network region.

### 3.2 The VGA Model

A Genetic Algorithm (GA) is a search based optimisation algorithm based on *Darwin's* principle of *survival of the fittest.* It combines the exploitation of past results with the new results. It can also mimic some of the flairs of human search with a structured yet randomized information exchange as well as a selection of the fittest technique.

A normal, unmodified genetic algorithm, with linear encoding of the variable and a simple two point crossover, when applied to the problem, was found to produce children which did not represent members of the solution set. Thus, in order to maintain the validity of the population after a crossover operation, the genetic operators were suitably modified. The modified version of the genetic algorithm used, namely Fully Independent Vector GA, preserves the validity of the solution. The variables are mapped as a 2D matrix represented in Figure 2 (a) and are described in Section 3.2 (a). The crossover is restrictive in the sense, at best a variation of one chromosome is allowed during cross over.

This also results in a much faster convergence. The only trade off in this scheme is the increase in complexity of the crossover operation both in terms of understanding and computational. However, the increase in computational complexity of the crossover is offset by the faster convergence rate and results in an overall decrease in time taken to find the solution.

A population is characterized by a set of chromosomes, which represent feasible solutions of the present step. We assume that each member of the population represents a borrowing decision executed by borrowing a single channel or more from one or more neighbours. Any $i^{th}$ member of the population carries four different information for cell number **i** and neighbours of cell **i**. A member of the population representing a cell and its neighbour cells is shown in Figure 2.

| 6 | 0 | 7  | 17 | 0 |
|---|---|----|----|---|
|   | 6 | 0  | 7  | 1 |
|   | 0 | 0  | 10 | 2 |
|   | 3 | 0  | 8  | 2 |
|   | 4 | 0  | 10 | 0 |
|   | 2 | 0  | 8  | 0 |
|   | 4 | 0  | 10 | 0 |

**Figure 2(a)**

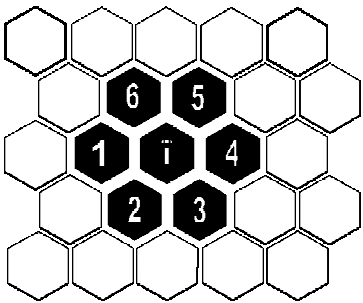

**Figure 2(b)**

(a) Characteristics of a single member of the population
(b) Cell i and its neighbouring cells in the network model

Figure 2(a) represents a single member of the population. The first row stands for the cell **i** who is imploring for channel(s) to its neighbours for some duration. Subsequent rows are the information about the neighbours of cell **i** numbered in anti clock wise starting from left in our network model.

*a)    Encoding*

In our VGA approach the encoding mechanism is an integral discrete scheme where a member of the population is composed of 7 chromosomes and each chromosome has four genes. The four different fields are described as

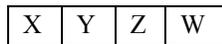

**Figure 3.** A gene structure

Where X, Y, Z and W represent number of *free channels,* number of *blocked hosts*, number of MHs present, and *cell type* respectively. Cells are of 5 types (i.e. 0 to 4). Cell type 0 has 6 neighbours and type 4 has 2 neighbours.

*b)    The VGA Structure*
The structure of VGA is as follows:

VGA( ) {
    Initialize population by seeding;
    Evaluate the population;
    While (termination condition not satisfied) {
        Perform fitness test;
        Select parents;
        Perform crossover;
        Perform mutation;
        Recombine population;
    }
}

*c)    Solution Space*

The solution space of the algorithm may be enumerated as thus:

Let each neighbour of cell **i** has a maximum of *n* borrowable channels. Each chromosome has 7 number of genes out of which one is used for data storage. Each gene may represent $5n^2$ number of values. Thus the solution space has an upper bound of $(5n^2)^6$. While a small value of *n* may indicate the feasibility of a brute force hill climbing method. Increase in *n* will quickly escalate the time complexity of a brute force approach in an exponential manner.

A GA is said to be efficient, if it converges to a solution within $\frac{r^n}{m} - 1$ number of iterations. Where m, and $r^n$ represent the size of the population, and search space respectively. In a practical situation when the number of channels available to a BS exceeds 50, the numbers of possible solutions exceed $10^{24}$.

*d)    Fitness Calculation*

The fitness value is used to rank the chromosomes for testing their quality. Each chromosome which represents a valid solution in the search space is assigned a fitness value according to the predetermined factors. Pruning may be performed to remove infeasible or unacceptable solutions by assigning low values. The factors considered for fitness are of unequal weight. The fitness function applied in the proposed VGA may be abstracted as follows:

fitness (childGene) = Channel left after borrowing + (α x channels borrowed) – (β x cell population) – (μ x cell type) + γ

Where α, β, μ and γ are of the order $10^{-1}, 10^{-2}, 10^{-3}$ *and* $10^{-4}$ respectively. Thus solutions obtained can be graded and then recombined with the existing population. Constants of the same order can be modified in order to give precedence to different factors. The last factor γ is a random mutation factor which affects only the fitness of the child and does not affect the gene. This is incorporated to protect the information of the gene while being able to differentiate between different solutions also.

*e)    Seeding*

Seeding is the process of selecting the initial population for faster convergence of the GA. In our model, the seeding was performed by borrowing *k = 1, 2 ... n* where *n* is the number of channels required by the cell, from each neighbour where the neighbour has at least *n* channels. The fitness of each member of the initialized population is calculated and stored in a vector with corresponding indices.

*f)    Crossover*

The standard crossover methods such as random single point or two point crossover are not applicable in a micro-layer approach as this would lead to an infeasible solution. Thus a highly restrictive non-standard crossover technique is implemented with a single cut point where relevant data is kept safe.

Parent A

| 0 | 2 | - | - |
| 0 | 1 | - | - |
| 0 | 0 | - | - |
| [2 | 0 | - | -] |
| [2 | 0 | - | -] |
| 1 | 0 | - | - |
| 0 | 1 | - | - |

Parent B

| 0 | 2 | - | - |
| 0 | 1 | - | - |
| 0 | 0 | - | - |
| [3 | 0 | - | -] |
| [1 | 0 | - | -] |
| 1 | 0 | - | - |
| 0 | 1 | - | - |

**Figure 4** An example of crossover operation between Parent A and Parent B

The crossover operation defined in our VGA scheme utilises the data in the first two columns of a member of the population matrix. The fixed cut point is taken which divides the matrix into two portions, the first row and the rest of the matrix. For the 2$^{nd}$ half of the child matrix, the values for the parents are compared and the lower value is copied into the child. Then the copied values are compared with the original request to determine the magnitude of difference and this difference is subtracted from the original value of blocked host for the cell requesting for channels. The crossover operation produces only one child. The parents are removed from the population and the child is re-inserted into the population after the child succeeded in its fitness test. An example of the said crossover operation is illustrated in Figure 4. Values marked by "-" are ignored during the crossover operation.

Child

| 0 | (1) | - | - |
| 0 | 1 | - | - |
| 0 | 0 | - | - |
| [2 | 0 | - | -] |
| [1 | 0 | - | -] |
| 1 | 0 | - | - |
| 0 | 1 | - | - |

Original Request

| 0 | (3) | - | - |
| 0 | 1 | - | - |
| 0 | 0 | - | - |
| 3 | 0 | - | - |
| 2 | 0 | - | - |
| 1 | 0 | - | - |
| 0 | 1 | - | - |

**Figure 5(a)**       **Figure 5(b)**
(a) A child (solution) produced from Parent A and Parent B
(b) An original request

g)   *Sample Input Output of VGA*

Input

| 0 | (5) | 0 | 0 |
| 1 | 0 | 9 | 1 |
| 6 | 0 | 4 | 2 |
| 0 | 1 | 11 | 2 |
| 0 | 3 | 13 | 0 |
| 5 | 0 | 5 | 0 |
| 0 | 4 | 14 | 0 |

Output

| 0 | (0) | 0 | 0 |
| 1 | 0 | 9 | 1 |
| 3 | 0 | 4 | 2 |
| 0 | 1 | 11 | 2 |
| 0 | 3 | 13 | 0 |
| 3 | 0 | 5 | 0 |
| 0 | 4 | 14 | 0 |

**Figure 5.** A sample input and output of VGA

A request for five channels is made and the corresponding information of the neighbours is passed to the VGA and the return is displayed as shown in Figure 6.

h)   *GA Termination Conditions*

The proposed VGA terminates if any one of the following conditions are reached:
1.   the size of the population reduces to 1
2.   the fittest chromosome doesn't require any channel
3.   the size of the population is 2 and both members represent the same solution.

**4. EXPERIMENTAL STUDY**

In this section we present the results of the simulation studies, carried out by us.

*Simulation Environment*

As a benchmark the standard FCA has been used. The micro-level VGA borrowing scheme is compared against the FCA and Greedy Borrowing Heuristic scheme. The standard test conditions are described as follows:

A uniform host distribution indicates an equal number of hosts present in each cell. Maximum supported hosts is the total number of host which can be supported by the entire network region under ideal conditions, i.e. *Number-of-cells * Channels-per-Cell.*

**Table 1.** Simulation Parameters

| Number of Cells | 25 |
|---|---|
| Channel Per Cell | 10 |
| Iterations | 20 |
| Host Distribution | Uniform |
| Number of Hosts | 200 to 250 |
| Max. Supported Host | 250 |

**Table 2.** Simulation Results

|  | Avg. Number of Blocked Hosts | | | | | |
|---|---|---|---|---|---|---|
| Number of Hosts | 200 | 210 | 220 | 230 | 240 | 250 |
| FCA (Benchmark) | 8.7 | 12.15 | 17.85 | 19.6 | 24.6 | 27.8 |
| SB (Greedy Borrowing) | 0.45 | 1.6 | 4 | 4.8 | 9.55 | 12.5 |
| VGA (Proposed) | 0 | 0.9 | 3.25 | 2 | 5.3 | 3.75 |

*Comparison with Related Works*

The performance of our proposed VGA was analysed over a varying degree of iterations as well as a varied number of hosts. The model was simulated 10 times for 200, 210, 220, 230, 240 and 250 number of MHs separately, and then the average value is taken at each iteration point which is represented in Table 2. In [6] four number of simulations were carried out: two simulations for FCA scheme and other two are for borrowing scheme with 64, 128 and 256 number of MHs. However we have considered 200, 210, 220, 230, 240, and 250 number of MHs with different number of iterations. The results show that our proposed model MiCi outperformed both the FCA and the SB Greedy Heuristic for all the test cases. The percentage of blocking at maximum load was seen to be around 3.75% for the proposed model where it is 7% (approx.) in [6] and 5.5% in [7].

The robustness of MiCi is attributed to the incorporation of VGA, which keeps the variable separate through out the number of generations and also carries out the seeding at the beginning of execution of the model. The novelty of the proposed scheme is to support high load and scaling to a large network in such a manner that its overall performance is better both on a short and long term basis.

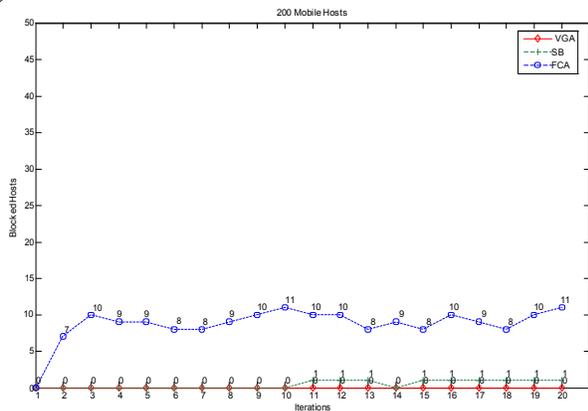

**Figure 6.** Blocking performance with 200 MHs

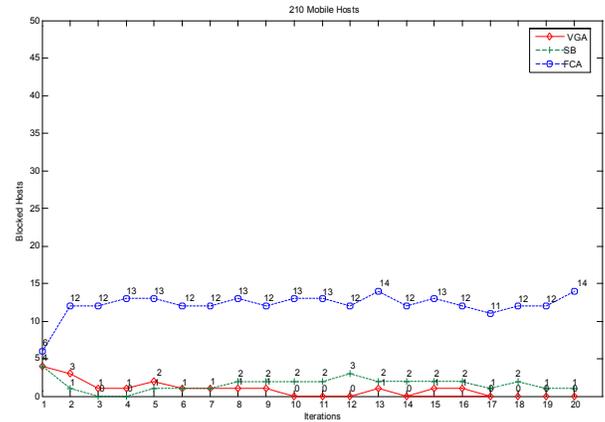

**Figure 7.** Comparison with 210 MHs

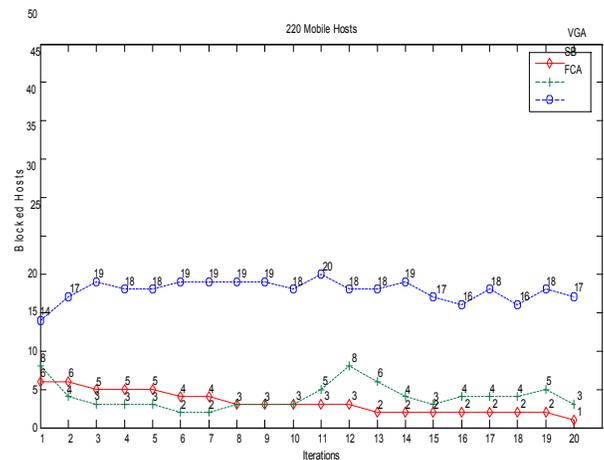

**Figure 8.** Blocking performance with 220 MHs

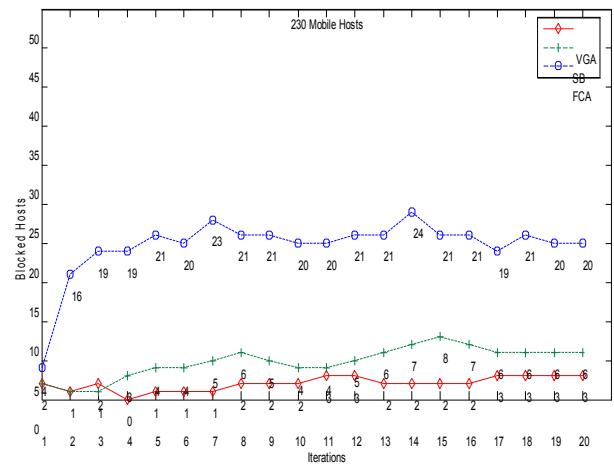

**Figure 9.** Blocking performance with 230 MHs

*MiCi: A Novel Micro-Level Temporal…*

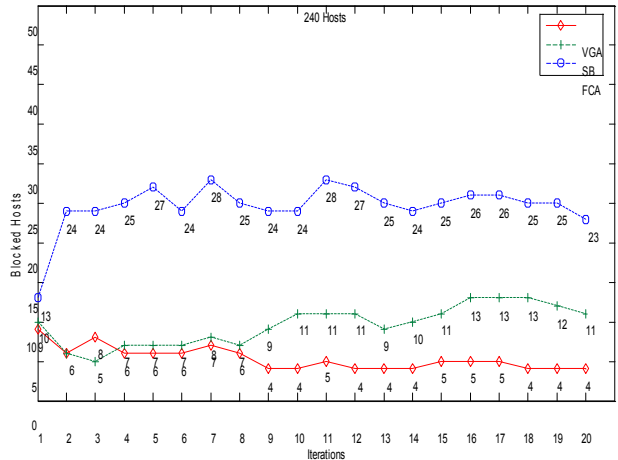

**Figure 10.** Blocking performance with 240 MHs

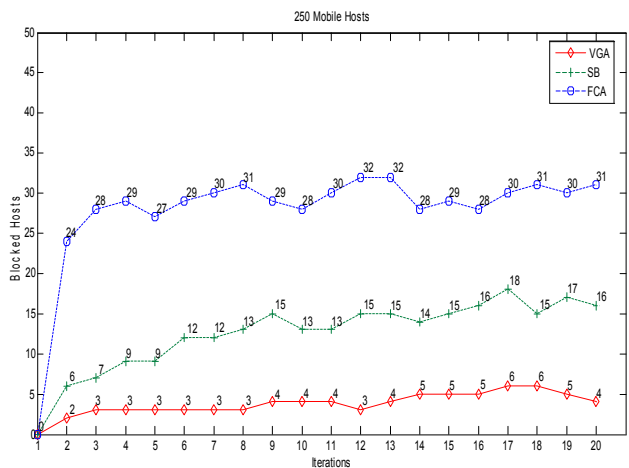

**Figure 11.** Blocking performance with 250 MHs

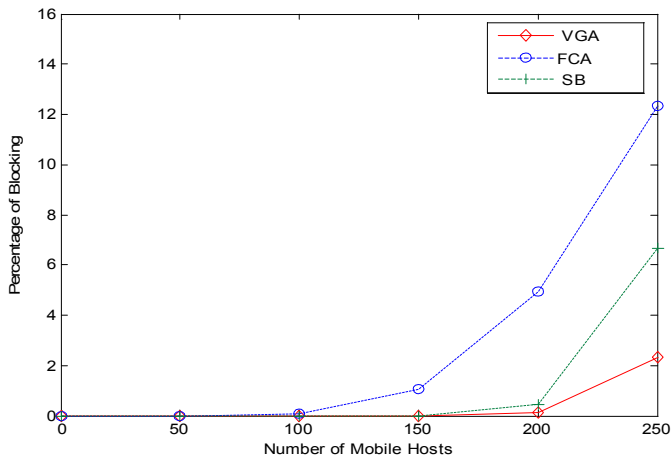

**Figure 12.** Average blocking percentage with varying number of MHs

## 5. CONCLUSION

We have proposed a Micro-layer temporal Channel imploring (MiCi) mechanism for Mobile Hosts (MHs) in wireless network which incorporates a Vector GA (VGA) for effective utilization of available channels in the network. Simulation studies were carried out with a network of rectangular shape containing 25 cells. The adoption of micro-layer approach and the VGA carries problem specific knowledge and future considerations for the stability of the network. The method essentially tries to borrow channels from neighbours without depleting channels of any specific neighbour completely. Further considerations made include the location of the cell from where the channel is being borrowed and future state of that cell. The cells are populated with such hosts and they are free to roam around but not free to leave the region. When a host leaves a cell and enters another cell, a hand-off is initiated and if required the VGA is invoked by MiCi. The VGA is required for MiCi only if the cell into which the host moves into does not have any free channels.

Our simulation results show that, compared to FCA and Greedy borrowing heuristics, they outperform the results of macro layer implementations in [6], [7] and [11] with respect to number of blocked hosts. As a future work, we plan to consider Support Vector Machine (SVM) with VGA for channel allocation.

**REFERENCE**

[1] Katzela, and M. Naghshineh, "Channel Assignment Schemes for Cellular Mobile Telecommunication Systems; A Comprehensive Survey," *IEEE Personal Communication,* Vol 31, No. 3, pp. 10-31, June 1996.

[2] W. K. Hale, "Frequency assignment: theory and assignment," Proc. IEEE, vol. 68, pp. 1497-1514, 1980.

[3] W. Wan, and W. Wong, "A heuristic algorithm for channel allocation of multi-rate data in hybrid TDMA/FDMA digital cellular systems," Proc. of IEEE Int. Symposium on Personal, Indoor and Mobile Radio Communications, Vol. 2, pp. 853-858, 1998.

[4] S. Hwang, J. Park, Y. S. Jang, and H. Cho, "A Heuristic Method for Channel Allocation and Scheduling in an OFDMA System," ETRI Journal, Vol. 30, No. 5, pp. 741-743, October 2008.

[5] T. M. Ko, "A frequency Selective Insertion Strategy for Fixed Channel Assignment," Proc. of 5th IEEE Int. Symposium on Personal, Indoor and Mobile Radio Communications, pp. 311-314, Sept. 1994.

[6] Albert Y.Zomaya, and Michael Wright, "Observations on Using Genetic-Algorithms for Channel Allocation in Mobile Computing ", IEEE Transactions on Parallel and Distributed Systems, Vol. 13, No. 9, September 2002

[7] Somnath Sinha Maha Patra, Kousik Roy, Sartak Banarjee, and Deo Prakash vidyarthi, "Improved Genetic Algorithm for Channel Allocation with Channel Borrowing in Mobile Computing", IEEE Transaction on Mobile Computing, Vol. 5, No. 7, July 2006

[8] Yao-Tien Wang, Chung Ming Ou, and Hsiang-Fu Yu, "Hierarchial Genetic Algorithms for Channel Allocation in Wireless Networks", 2008 IEEE Asia-Pacific Services Computing Conference

[9] Sasthi C. Ghosh, Bhabani P. Sinha, and Nabanita Das, "Channel Assignment Using Genetic Algorithm Based on Geometric Symmetry", IEEE Transactions on Vehicular technology, Vol. 52, No. 4, July 2003

[10] Marcos A. C. Lima, Aluizio F.R. Arauzo, and Amilkar C. Cesar, "Adaptive Genetic Algorithms for Dynamic Channel Assignment in Mobile Cellular Communication Systems", IEEE Transactions on Vehicular technology, Vol. 56, No. 5, September 2007

[11] Lutfi Mohammed Omer Khanbary, and Deo Prakash Vidyarthi, "A GA-Based Effective Fault-Tolerant Model for Channel Allocation in Mobile Computing", IEEE Transactions on vehicular Technology, Vol. 57, No. 3, November 2008

[12] Hassan M. Elkamchouchi, Hassan M. Elragal, and Mina A. Makar, "Channel Assignment for Cellular Radio Using Particle Swarm Optimization", The 23rd National Radio Science Conference (NRSC 2006) March 14-16, 2006.

[13] Dietmar Kunz, "Channel Assignment for Cellular Radio Using Neural Networks", IEEE Transactions on Vehicular technology, Vol. 40, No. 1, February 1991

[14] Manuel Duque-Anton, Dictmar Kunz, and Bernhard Ruber, "Channel Assignment for Cellular Radio Using Simulated Annealing", IEEE Transactions on Vehicular technology, Vol. 42, No. 1, February 1993.

[15] Justin C.-I. Chuang, "Performance Issues and Algorithms for Dynamic channel Assignment", IEEE Journal on Selected Areas in Communications, Vol. 11, No. 6, August 1993.

**BIOGRAPHY**

**Snehasish Kumar** has received his Bachelors degree in Computer Science and Engineering from Silicon Institute of Technology under Biju Pattanaik University of Technology, Orissa, India in 2010. He is about to pursue his Masters Degree from Simon Fraser University, Vancouver, Canada in Computer Science. His research interests lie in the fields of wireless and mobile networks with the application of machine learning and optimization techniques based on soft computing to these fields. He is a student member IEEE.

**Satyananda Champati Rai** received M. Tech. degree in Computer Science from Utkal University in 2001. Presently he is working as an Asst. Professor in the Department of Computer Science and Engineering, Silicon institute of Technology, Bhubaneswar, India and pursuing his Ph.D. work. His research interests include QoS for wireless network, mobile ad hoc and sensor networks.

**Rajib Mall** received B.Tech., M.Tech. and Ph.D. degrees from the Indian Institute of Science, Bangalore in 1983, 1988 and 1992 respectively. During 1992-94 he worked for Motorola India Ltd. He joined the Department of Computer Science and Engineering, Indian Institute of Technology, Kharagpur in 1994 as a faculty member, and is currently a Professor. His current research interests include design and software engineering issues in large systems and those in real-time embedded systems.

**Sateesh K. Pradhan** has obtained his Ph.D. degree in Computer Science from Berhampur University, India in 1999. He joined Berhampur University, as Assistant Professor in the year 1987 and promoted to Associate Professor in 1999. He was Head of the Department of Computer Science, Utkal University, India during 2001-2003. He was the Secretary and Vice-President of Orissa IT Society from 2003-2005 and 2005-2007 respectively and was the organizing Chair of the International Conference on Information Technology – 2005 (ICIT – 2005). At present he is on leave from Utkal University and working with the Department of Computer Engineering, King Khalid University, KSA. His research interests include Neuron based Parallel Algorithms, Computer Architecture, Ad-hoc Network and Computer Forensic.